\documentclass[12pt]{iopart}

\usepackage{graphicx}

\def\ii{\'{\char'20}}
\def\beq{\begin{equation}}
\def\eeq{\end{equation}}
\def\bea{\begin{eqnarray}}
\def\eea{\end{eqnarray}}
\def\dGG{\dot{G}/G}
\def\fdGG{\frac{\dot{G}}{G}}
\def\daa{\dot{\alpha}/\alpha}
\def\fdaa{\frac{\dot{\alpha}}{\alpha}}
\def\dRR{\dot{R}/R}
\def\fdRR{\frac{\dot{R}}{R}}
\def\lapprox{\hbox{\lower .8ex\hbox{$\,\buildrel < \over\sim\,$}}}
\def\gapprox{\hbox{\lower .8ex\hbox{$\,\buildrel > \over\sim\,$}}}

\begin{document}
\jl{6}

\title[Time  variation of $G$ and $\alpha$]{Time  variation of $G$ and
$\alpha$ within models with extra dimensions}

\author{P.  Lor\'en--Aguilar\dag,  E.  Garc{\ii}a--Berro\dag\ddag,  J.
	Isern\ddag$\|$,   and  Yu.A.~   Kubyshin\P\S}

\address{\dag\   Departament  de  F{\ii}sica   Aplicada,   Universitat
	 Polit\`ecnica de Catalunya, c/Jordi Girona, 1-3, M\`odul B-4,
	 Campus Nord, 08034 Barcelona, Spain}
\address{\ddag\ Institut d'Estudis Espacials de Catalunya, Ed.  Nexus,
	 c/Gran Capit\`a 2, 08034 Barcelona, Spain}
	 \address{$\|$\ Institut de Ci\`encies de l'Espai (CSIC)}
\address{\P\  Departament  de  Matem\`atica  Aplicada IV,  Universitat
	 Polit\`ecnica de Catalunya, c/Jordi Girona, 1-3, M\`odul C-3,
	 Campus Nord, 08034 Barcelona, Spain}
\address{\S\  On leave  of  absence  from  the  Institute  of  Nuclear
	Physics, Moscow State University, 119899 Moscow, Russia}

\begin{abstract}
We derive the formulae  for the time  variation  of the  gravitational
``constant''  $G$ and of the fine structure  ``constant''  $\alpha$ in
various  models with extra  dimensions  and analyze their  consistency
with the available observational data for distant supernovae.  We find
that  the  reported  variation  of  $\alpha$  translates  into a small
variation  of  $G$  that  makes  distant  supernovae  to  appear  {\sl
brighter},  in  contradiction  with  recent  observations  of high $z$
supernovae.  The significance of these results within the framework of
some cosmological scenarios is also discussed.  We find, however, that
the magnitude of the effect is not large enough to safely  discard the
models with extra dimensions studied here.
\end{abstract}


\section{Introduction}
\label{Intro}

One  of  the  most  challenging   issues  of  modern  physics  is  the
quantization   of   the   gravitational   interaction.   Within   this
theoretical   framework  it  is  worth   noticing  that  although  the
controversial  issue of the  variation  of the  fundamental  couplings
dates back to the large-number  hypothesis of Dirac  \cite{Dirac},  it
has  recently   become  a  subject  of  intensive   experimental   and
theoretical  studies --- see  \cite{Uzan02}  for an excellent  review.
Modern  theories,  like the  string/M-theory  or brane  models, do not
necessarily require a variation of the fundamental  constants but they
provide a natural and  self-consistent  framework for such variations.
Most of such theories assume the existence of additional dimensions of
the space-time and, therefore, contain a built-in  mechanism  allowing
for the time variation of the observed  couplings in four  dimensions.
This   feature   is   rather   easy   to    understand.   Within   the
multidimensional   approach  the  interactions   are  described  by  a
fundamental   theory   formulated   in  $4+d$   dimensions,   and  the
conventional  four-dimensional  theory  appears  as a  result  of  the
dimensional reduction.  Couplings in four dimensions are determined by
a set of a few constants of the  multidimensional  theory and the size
$R$ of the space of extra dimensions.  The multidimensional  constants
are assumed to be genuinely fundamental and, consequently, do not vary
with time.  On the other hand, in the  cosmological  or  astrophysical
context it is natural to assume  that $R$ varies  with time, very much
in the same way as the scale  factor  of the  three-dimensional  space
does.  This leads to the time variation of parameters of the effective
four-dimensional  theory, like the gravitational constant $G$ and fine
structure  constant  $\alpha$  \cite{var-Xdim}--\cite{KOP}.  Moreover,
since their time  dependences  are given by the same scale  factor $R$
these  variations may turn out to be correlated,  and it is natural to
wonder  what  kind  of  relation  is  predicted  in   multidimensional
scenarios.  The effect of these  correlations for various couplings in
the  framework  of  theories  of  unification  has been  discussed  in
\cite{LSS}.  It has also been suggested  that these  correlations  may
unveil the mechanism of the time  variation of the  couplings  and, in
this way, give insight into the underlying  fundamental theory.  It is
important to emphasize that a variation of the  fundamental  couplings
may lead to deeper  consequences  or effects in the theory of particle
interactions.  In  particular,  it has been shown that a variation  of
fundamental couplings can be intrinsically  related to the Lorentz and
CPT violation \cite{KoPo}.

There have been several  attempts to measure the rate of  variation of
the  fundamental  constants  but many of them have yielded  just upper
bounds on the  absolute  value of the rate of  change.  However,  very
recently,  the rate of variation of the fine  structure  constant  has
been measured using high  resolution  spectroscopy  of QSO  absorption
systems.  To be specific a non-zero  detection of the variation of the
fine structure constant, $\Delta\alpha/\alpha\sim  -10^{-5}$ at $z\sim
1.5$, has been reported  \cite{Mur00}.  Although  this result is still
the subject of strong debate it is worth studying its theoretical  and
observational  consequences.  In addition,  the  observations  of high
redshift  $(z>0.1)$  Type Ia  supernovae  \cite{SCP,  HzSST}  and  the
analysis of the CMB strongly  suggest a flat,  $\Omega_R=0$,  Universe
\cite{omegar},  with  a  mass  density   $\Omega_M\simeq  0.3$  and  a
non-vanishing cosmological constant  $\Omega_\Lambda\simeq 0.7$.  This
result, in turn, motivated a considerable amount of papers looking for
viable  alternatives.  Amongst these  alternatives  the variability of
the  gravitational  constant  was  one of  the  proposed  alternatives
\cite{Amen99}   to  reconcile  the   observational   data  of  distant
supernovae with an open $\Omega_\Lambda=0$ Universe.

As it has  been  argued  in  \cite{BDD},  models  in  which  the  time
variation of couplings  --- and in  particular  of the fine  structure
constant --- is generated  by the  dynamics of a  cosmological  scalar
field  face  serious   difficulties.  Namely,  using  rather   general
arguments,   these   authors   show  that  in  order  to  explain  the
observations of \cite{Mur00} such models require an extremely  precise
fine-tuning  that  cannot be  explained  by any known  mechanism.  The
problem is  essentially  the huge  back-reaction  produced  by varying
couplings on the vacuum energy.  This  difficulty  could be intimately
related to the long-standing  cosmological  constant  problem.  Hence,
its  satisfactory  solution could also provide a mechanism to suppress
the enormous  variation of the vacuum energy due to the time variation
of  $\alpha$,  but this is  beyond  the  scope of this  paper.  Having
adopted this point of view, we do not discuss  this issue  furthermore
in the  present  work.  Instead,  the aim of the our work is to derive
the relations  between $\dGG$ and $\daa$ in various  models in $(4+d)$
space-time  dimensions  and to  confront  them with the  observational
data, namely to compare the detected  variation of the fine  structure
constant \cite{Mur00} with the existing bounds on the variation of $G$
at  cosmological  distances.  The  models  we are  going  to  consider
include the classical  Kaluza-Klein models --- see  \cite{KK-rev,Duff}
for  reviews  on  the  subject   ---  models   with   multidimensional
gravitational  and  Yang-Mills  fields   \cite{CSDR,KMRV89},  and  the
Randall-Sundrum  model with two branes and gauge and fermionic  fields
propagating  in  the  bulk   \cite{RS1}--\cite{DHR01}.  The  paper  is
organized as follows.  In Section \ref{KKmodels} formulae for the time
variation  of $G$ and  $\alpha$  in various  models  are  derived.  In
Section   \ref{observ}  we  use  our   theoretical   results  and  the
experimental  value for $\daa$  \cite{Mur00}  to obtain an estimate of
the rate $\dGG$, which is then confronted  with the data obtained from
distant  Type Ia  Supernovae  \cite{SCP,HzSST}.  Finally,  in  Section
\ref{conclusions}  conclusions and some  discussion of the results are
presented.

\section{Time variation of couplings in multidimensional theories}
\label{KKmodels}

In this  section we consider  theories with extra  spatial dimensions.
For the sake of simplicity we assume that the geometry of the space of
additional dimensions is  described by just one scale  factor $R$.  In
most of the  cases the generalization to geometries  with a few scales
is straightforward  and does not bring any  qualitatively new features
to the effect  we are going to study.   Moreover, the theories studied
here are  assumed to be part  of a cosmological  scenario with varying
$R$.  The specific form of  the function $R(t)$ depends on the details
of the  scenario --- see,  for instance, \cite{KoTu}.  Here  we simply
assume  that the  evolution of  the multidimensional  Universe  at the
cosmological  scale  is  described  by some  scenario  which  predicts
certain background metric with varying scale $R(t)$.  Another possible
class of scenarios  would be the ones postulating  a first order phase
transition in the  Universe.  However, we will not  consider this last
possibility here  and we limit ourselves  to multidimensional theories
describing the  time variation of  $G$ and $\alpha$ which  fulfill the
following  conditions: (1) the  four-dimensional effective  theory ---
that is,  the dimensionally reduced theory ---  includes both Einstein
gravity and Maxwell electrodynamics; and (2) the time variation of the
scale factor $R(t)$ of the space of extra dimensions leads to the time
variation  of  both  the  gravitational  constant  $G$  and  the  fine
structure constant $\alpha$.  To  calculate $\dGG$ and $\daa$ we focus
only on  the gravitational and electromagnetic sectors  of the reduced
theory.  These time variations as  well as that of $R(t)$ are supposed
to be slow enough in  comparison with the phenomena described by these
sectors ---  namely, the electromagnetic processes  at the microscopic
level.  Our analysis  will be rather general and will  not rely on any
particular function $R(t)$.

\subsection{Kaluza-Klein theories}

In two pioneering papers \cite{KK} T.  Kaluza and O.  Klein formulated
the  essential  elements  of  the  multidimensional  approach  to  the
description  of  fundamental  interactions  which was later called the
Kaluza-Klein  approach.  They considered the pure Einstein  gravity in
five-dimensional  space time $M^{4}  \times  S^{1}$  described  by the
multidimensional  metric  tensor  $\hat{g}_{MN}$  and showed  that the
sector of zero modes of the dimensionally  reduced theory includes the
four-dimensional  gravity  and  Maxwell  theory.  Here  $M^{4}$ is the
four-dimensional  Minkowski  space-time,  and  $S^{1}$ is the  circle.
Later this  construction  was  generalized  \cite{KK1} to more general
compact spaces of extra dimensions.  In these cases the reduced theory
contains  Einstein gravity and Yang-Mills  fields with the gauge group
determined by the isometry group of the space of extra dimensions.  In
the standard setting the  multidimensional  Kaluza-Klein theory is the
pure Einstein theory on $M_{4} \times K_{d}$ with the action given by
\beq
S = \int d^{4+d} \hat{x} \sqrt{-\hat{g}} \frac{1}{16 \pi G_{(4+d)}} 
{\cal R}^{(4+d)}.
   \label{D-grav}
\eeq
Here $M_{4}$ is a (curved)  four-dimensional  space-time, $K_{d}$ is a
compact manifold of extra dimensions,  $\hat{g} = \det  \hat{g}_{MN}$,
($M,N  =  0,1,2,  \ldots,  3+d$),  ${\cal  R}^{(4+d)}$  is the  scalar
curvature   in  $M_{4}   \times   K_{d}$,   and   $G_{(4+d)}$  is  the
multidimensional  gravitational constant, which is assumed to be truly
constant and does not depend on time.  According to the  procedure  of
dimensional  reduction,  to  obtain  the  four-dimensional   effective
theory, firstly the $\mu\nu$-components  $(\mu,\nu=0, 1, 2, 3)$ of the
metric tensor are identified as the  four-dimensional  metric  tensor.
Certain combinations of the rest of the components, $\hat{g}_{\mu m}$,
$\hat{g}_{n  \nu}$  and  $\hat{g}_{mn}$  with  $m,n=4,\ldots,3+d$  are
identified  as gauge  field  multiplets  $A_{\mu}$  and scalar  fields
$\phi_{mn}$.  Secondly,  the mode  expansion  of all  these  fields is
performed --- see, for example,  \cite{Duff}.  The coefficients of the
expansion   depend  only  on   $x^{\mu}$   and  are   interpreted   as
four-dimensional  fields.  In general  there is an infinite  number of
them but here we are  interested  in the sector  containing  only zero
modes of the mode expansion.  Its action is given by
\beq
S_{0} = \int d^{4}x \left[  \frac{1}{16 \pi G(t)} {\cal R}^{(4)} +
\sum_{i} \frac{1}{4 g_{i}(t)^{2}} Tr F^{(i)}_{\mu \nu} F^{(i) \mu \nu} 
\right],
\label{DR-action}
\eeq   
where $G(t) \equiv G_{(4)}(t)$ is the  four-dimensional  gravitational
constant.  The parameters  $g_{i}(t)$ are the gauge couplings, and the
index  $i$  labels  the  simple  subgroups  of the  gauge  group.  The
dimensional reduction of the initial Kaluza-Klein action $S$, given by
Eq.  (\ref{D-grav}),  yields  in  addition  to  $S_0$,  given  by  Eq.
(\ref{DR-action}),    terms   including    non-zero   modes   of   the
gravitational,  gauge and scalar fields as well as terms  proportional
to  $(\dRR)^2$.  The scalar  fields  usually  give  highly  non-linear
interaction terms and are coupled  non-minimally to the  gravitational
and gauge fields.  For the sake of  simplicity,  the scalar fields are
supposed  to be  frozen  out  and  their  contribution  is  neglected.
Identifying  the  gravitational  and gauge  couplings  from the action
$S_0$ for the zero  modes we  obtain  the  following  expressions  for
$G(t)$ and $g_i(t)^2$ in terms of $G_{(4+d)}$ and the radius $R(t)$ of
the space of extra dimensions:
\bea
   G(t) & = & \frac{G_{(4+d)}}{V_{d}(t)},  \label{GG}    \\
   g_{i}^{2}(t) & = & \tilde \kappa_{i} 
                      \frac{G_{(4+d)}}{R(t)^{2} V_{d}(t)},   
	    \label{gG}
\eea
where $V_{d}(t)  \sim R(t)^{d}$  is the volume  of the space  of extra
dimensions and $\tilde\kappa_{i}$ are coefficients which depend on the
isometry group of $K_{d}$.  Equations (\ref{GG}) and (\ref{gG}) should
be regarded as leading order  approximations. The terms omitted in the
reduced  action  will  give  sub-leading  corrections,  in  particular
through loop  effects. As it has  been said above, we  assume that the
dimensionally reduced  theory includes the  electrodynamics.  Then the
fine structure  constant $\alpha(t)$ is given by  a linear combination
of $g^{2}_{i}(t)$,  the specific relation  depending on the  model, in
particular  on the  gauge  group  and the  scheme  of the  spontaneous
symmetry breaking.  From Eq. (\ref{gG}) it follows that
\beq
\alpha(t) = \kappa_1  \frac{G_{(4+d)}}{R(t)^{2} V_{d}(t)}, \label{aG}
\eeq
where  $\kappa_1$ is some constant.  Therefore, the time  variation of
$G$  and  $\alpha$  is  determined  by  the  function   $R(t)$.  Since
$\dot{V}_{d}/V_d = d (\dRR)$, we get
\bea
  \fdGG & = & - d \fdRR,                 \label{dGG-KK}      \\
  \fdaa  & = & - (d+2) \fdRR.       \label{daa}
\eea
As a  consequence  the  time  variation  of  the  fine  structure  and
gravitational constants are related by
\beq
    \fdaa = \frac{d+2}{d} \fdGG .      \label{daG}
\eeq

\subsection{Einstein-Yang-Mills theories}

Consider a theory in  the $(4+d)$-dimensional space-time $M_{4} \times
K_{d}$ that includes gravity and the Yang-Mills field with the action
\beq
S = \int d^{4+d} \hat{x} \sqrt{-\hat{g}} \left[
  \frac{1}{16 \pi G_{(4+d)}} {\cal R}^{(4+d)} +
  \frac{1}{4 g_{(4+d)}^{2}} Tr \hat{F}_{MN} \hat{F}^{MN} \right] ,
   \label{D-EYM}
\eeq
where, as above,  $G_{(4+d)}$  is the  multidimensional  gravitational
constant, and  $g_{(4+d)}$  is the  multidimensional  gauge  coupling.
Both are supposed to be constant in time.  The  dimensionally  reduced
theory includes the Einstein  gravity and the  four-dimensional  gauge
fields with an action similar to that of  Eq.~(\ref{DR-action}),  and,
in addition,  scalar  fields with a quartic  potential.  The  explicit
form of the  dimensionally  reduced theory depends on the topology and
geometry  of the space of extra  dimensions  and the  multidimensional
gauge  group.  The  case of  $K_{d}$  being a  homogeneous  space  was
studied  in  detail  in  the  literature   \cite{CSDR}  ---  see  also
\cite{KMRV89}  for  reviews  on  the  subject.  The  four  dimensional
gravitational constant is given by
\bea
	G(t) & = & \frac{G_{(4+d)}}{V_{d}(t)},     \label{GG-1}
\eea
We assume that the gauge part of the initial multidimensional model is
such that  its dimensional reduction  gives the bosonic sector  of the
electroweak Glashow-Salam-Weimberg model  in four dimensions. Examples
of this  kind can be found  in \cite{KMRV89}. Then  the fine structure
constant  in the  reduced theory  is related  to  the multidimensional
gauge coupling $g_{(4+d)}$ by
\bea
	\alpha(t) & = & \kappa_2 \frac{g^2_{(4+d)}}{V_{d}(t)}.  
	\label{aG-1}
\eea
where $\kappa_2$ is some constant factor.  From these  expressions the
following relation between the time variations of $G$ and $\alpha$ can
be easily obtained:
\beq
    \fdaa = \fdGG .      \label{daG-1}
\eeq
We would  like to  mention here  that the same  realtion appears  in a
ten-dimensional  model obtained  as a  low  energy limit  of a  string
theory \cite{Mae88}.

\subsection{Randall-Sundrum model}

A different  multidimensional  setting motivated by  string/M-theories
was proposed and studied in  \cite{RS1}.  The model is  formulated  in
the  five-dimensional  space-time  $M^{4}  \times  K_1$ with the fifth
dimension compactified to the orbifold $K_1=S^{1}/Z_{2}$ of radius $R$
--- see \cite{RS-review-Ant} and  \cite{RS-review-Rub}  for reviews on
the subject.  In the initial version of the Randall-Sundrum model with
two branes located at the fixed points of the orbifold,  known also as
the RS1 model, only gravity propagates in the  five-dimensional  bulk.
The   background   metric   solution   is  given  by  \beq   ds^{2}  =
e^{-2kR(|\phi|-\pi)}   \eta_{\mu   \nu}  dx^{\mu}   dx^{\nu}  +  R^{2}
d\phi^{2} \label{RS-metric} \eeq  \cite{RS1,DDG,BKSV02},  where $\phi$
is  the   coordinate   of  the   orbifold   $(0\le   \phi  \le  \pi)$,
$\eta_{\mu\nu}$ is the  four-dimensional  Minkowski metric tensor, and
$k>0$ is a parameter of the  dimension  of mass, its value being fixed
by the brane tensions.  Our three-dimensional space is identified with
the brane with  negative  tension  at  $\phi=\pi$.  The  fields of the
Standard  Model are assumed to be localized at this brane.  Due to the
warp factor in the background metric,  Eq.~(\ref{RS-metric}),  the RS1
model  provides  an  elegant  geometrical  solution  to the  hierarchy
problem   \cite{RS1}.  In  addition  it  opens  new   phenomenological
possibilities  like  the  observation  of  higher-dimensional  gravity
effects in the current or future collider experiments \cite{Kub01}.

The reduction formula expressing the  four-dimensional  Planck mass in
terms of the  fundamental  (five-dimensional)  mass  scale  $M=(16 \pi
\hat{G}_{(5)})^{-1/3}  \sim k$ is derived \cite{BKSV02} using a metric
given by  Eq.~(\ref{RS-metric}),  which  corresponds  to the  Galilean
coordinates on the physical brane at $\phi=\pi$.  One gets
\beq
M_{Pl}^{2} = \frac{M^{3}}{k} \left[ e^{2\pi k R} - 1 \right] 
\label{RS-RF}
\eeq
(see also  \cite{RS-review-Rub}).  To generate  the correct  hierarchy
between the Planck scale and the  TeV-scale  the product $k R$ must be
$k R \approx 11 \div 12$.  As before, let us now  consider  this model
as a part of a more general cosmological  scenario with a slow varying
scale factor $R(t)$.  The background solution  (\ref{RS-metric})  must
be modified accordingly, in particular its four-dimensional part takes
the form of the  Robertson-Walker  metric with the scale factor $a(t)$
multiplied by the conformal warp factor  including  $R(t)$.  Models of
this type have been extensively studied in the literature --- see, for
example,  \cite{RS-cosmology}.  It  is  natural  to  expect  that  the
reduction  formula in the first  approximation  remains the same as in
Eq.~(\ref{RS-RF})  but with  $R=R(t)$.  From this  equation it readily
follows that
\beq 
G (t)=  \frac{k}{16  \pi  M^{3}}  \frac{1}{e^{2k\pi  R(t)} -1} 
\approx  \frac{k}{16 \pi M^{3}} e^{-2k\pi R(t)}.
\label{GM-RS}  
\eeq 
The time variation of the Newton constant is then given by
\beq
 \fdGG = - 2 \pi k R(t) \fdRR {1\over { 1 - e^{-2k \pi R(t)} }} 
 \approx  - 2 \pi k R(t) \fdRR           \label{dGG-RS}
\eeq
However,  since the fields of the Standard  Model are localized on the
brane and do not depend on $R$ the RS1 model does not contain a simple
mechanism  for  the  variation  of the  fine  structure  constant.  To
describe  this  effect one has to  consider  bulk gauge and,  perhaps,
fermionic  fields.  Such  models  have  been  studied  in a number  of
papers, see for example  \cite{DHR99,DHR01}.  We assume that,  similar
to the case of the Einstein-Yang-Mills theories studied in \S 2.2, the
bulk gauge fields yield the gauge (and  perhaps the scalar)  sector of
the  Standard  Model on the brane.  The  electromagnetic  $U(1)$-field
appears in a usual way as a part of this sector after the  spontaneous
symmetry  breaking.  Similar to  Eq.~(\ref{aG-1}),  the fine structure
constant  turns  out  to be  related  to  the  multidimensional  gauge
coupling $g_{(5)}$ by
\beq
    \alpha(t)=\kappa_3 \frac{g^{2}_{(5)}}{R(t)},   \label{aa-RS}
\eeq
where  $\kappa_3$  is  some  constant.  From  Eqs.~(\ref{dGG-RS})  and
(\ref{aa-RS}) we obtain the following relation:
\beq
\fdaa = \frac{1}{2\pi k R(t)} \fdGG (1 - e^{-2k \pi R(t)}) \approx
\frac{1}{2\pi k R(t)} \fdGG.       \label{daG-RS}
\eeq
We would  like to  mention at this  point that  as it was  observed in
\cite{KOP}  the   effect  of  the  time  variation   of  couplings  in
brane-world  scenarios is  closely related  to the  resolution  of the
hierarchy problem.

\subsection{General remarks} 

Let us make a few remarks.  Expressions (\ref{aG}),  (\ref{aG-1}), and
(\ref{aa-RS}) are classical, or tree-level relations.  They define the
fine structure constant $\alpha (M_R;t)$ at the scale $M_R=R^{-1}(t)$.
We  have  found   that   $\dot{\alpha}(M_R;t)/   \alpha(M_R;t)   =\rho
(\dot{R}/R$),  where $\rho$ is a constant.  In order to relate $\alpha
(M_R;t)$  to the  correspondig  value  $\alpha  (\mu  ;t)$ at some low
energy scale $\mu$, for example at the  electroweak  scale  $\mu=M_Z$,
one  should  take  quantum  corrections  into  account  by  using  the
renormalization  group  formulas  for running  couplings.  By standard
considerations one obtains a relation of the form
\[
\frac{1}{\alpha (\mu;t)} = \frac{1}{\alpha (M_R;t)} + A \ln (\mu/M_R),
\]
where  $A$  is  a  constant  of  order  one  \cite{ItZu}.  Though,  in
principle, the second term also  contributes to the time  variation of
$\alpha(\mu;t)$, in fact its variation is dominated by the first term,
$\alpha^{-1}(M_R;t)$.  The  reason  is quite  simple.  Let us take the
point of view that the scale  $\mu$ does not vary with time.  In other
words, the variation of the  dimensionless  quantity  $\mu/M_R(t)$  is
determined by $R(t)$.  Then by a  straightforward  time  derivation of
the previous expression one obtains:
\bea
\frac{\dot\alpha(\mu,t)}{\alpha(\mu;t)}=\rho\fdRR-
\alpha(\mu;t)A\fdRR\Big[1+\ln\mu R(t)\Big].
\label{mu-scale}
\eea  
The first term in this  expression is the time variation calculated in
\S 2.1, 2.2  and 2.3 for three specific classes  of models; the second
term is of order $\cal{O}(\alpha)$. Hence, it is sub-dominant and will
be neglected from now on. If the scale $\mu$ is defined in a different
way and appears  to be time dependent, the form of  the second term in
Eq.   (\ref{mu-scale})  may  change,   but  it   will  remain   to  be
sub-dominant.  A  similar analysis  was presented in  \cite{Mar84} and
\cite{LSS}.

In  order   to   summarize,   Eqs.~(\ref{daG}),   (\ref{daG-1}),   and
(\ref{daG-RS}) can be written in the following general form:
\beq
    \fdaa = \beta (R) \fdGG ,      \label{daG-gen}
\eeq
where
\beq
   \beta (R) = \left\{ \begin{array}{ll}
   \frac{d+2}{d}  & \mbox{for the Kaluza-Klein theories},  \\
   1 &   \mbox{for the Einstein-Yang-Mills theories},  \\
   \frac{1}{2 \pi kR(t)}   &   \mbox{for the Randall-Sundrum-type model}.
    \end{array}     \right.                       \label{beta-def}
\eeq
Note that $\daa \propto \dGG$, that the constant of proportionality is
positive, and that $\beta \sim 1$ in the case of the Kaluza-Klein  and
Einstein-Yang-Mills  theories, and $\beta \sim 10^{-2}$ in the case of
the Randall-Sundrum model.  We would like to emphasize that the result
given in  Eqs.~(\ref{daG-gen})  and  (\ref{beta-def}) is robust, since
does not depend on details  of the  models  and the  specific  form of
$R(t)$.

If the  dimensional reduction of  the multidimensional models  gives a
four-dimensional  model of  unification, then  an analysis  similar to
that  carried out previously  in subsections  2.1--2.3 gives  the time
variation of  a single coupling  constant $\alpha_{\rm GUT}$.  In this
case, Eq.  (\ref{mu-scale}) for  the coupling running is different and
relates the  three couplings of  the Standard Model  with $\alpha_{\rm
GUT}$.  The electromagnetic coupling at  the low energy scale $\mu$ is
calculated in  a standard way. Making  an analysis similar  to that of
\cite{LSS}  it can be  shown that  the time  variation in  the leading
approximation  is related  to $\dot\alpha_{\rm  GUT}/\alpha_{\rm GUT}$
by:

\[
\frac{\dot\alpha(\mu;t)}{\alpha(\mu;t)}\propto
\frac{\dot\alpha_{\rm GUT}}{\alpha_{\rm GUT}}
\]

\noindent where the proportionality constant is of order unity.

Finally,  let us also  mention  that  there is a class of models  with
branes and large extra  dimensions, the  Arkani-Hamed-Dimopoulos-Dvali
(ADD) models, which also provide a solution to the  hierarchy  problem
and predict new observable effects with massive gravitons  \cite{ADD}.
However,  similar to the case of the  RS1-model,  the ADD-model in its
standard version with only the gravitational  field propagating in the
bulk  does not  describe  the time  variation  of the  fine  structure
constant.

\section{Observational constraints on $\daa$ and $\dGG$}\label{astro}
\label{observ}

The detailed analysis of three distant $(z\sim 3.5)$ quasar absorption
line data  sets has provided for  the first time  direct evidence that
the fine  structure constant  $\alpha$ was {\sl  smaller} in  the past
\cite{Mur00}.   In  particular,  the   detection  of  a  variation  in
$\alpha$:
\beq
\frac{\Delta\alpha}{\alpha} \equiv 
\frac{\alpha (z) - \alpha_{0}}{\alpha_0}  
=(-0.66\pm  0.11)\times 10^{-5},   \label{delta-aa} 
\eeq 
where  $\alpha_{0} = \alpha (z_{0})$ is the present day value has been
reported.  It is of  importance  for  our  analysis  to  realize  that
$\Delta\alpha= \alpha(z)- \alpha_0 <0$.  We also emphasize that, since
the error bars are considerably  smaller than the reported value, this
is not just an upper  bound  but a direct  measurement  of the rate of
change of the fine  structure  constant.  Whether  this  detection  is
genuine  or is  affected  by  systematic  errors  is still a matter of
debate    \cite{Uzan02,Barr02}.   Using    Eqs.~(\ref{daG-gen})    and
(\ref{beta-def})  it turns out that,  provided that the  observational
determination  of  $\Delta\alpha/\alpha$  is correct, the variation of
$\dGG$ in the models studied in the previous section is {\sl positive}
and we  face a {\sl  smaller}  value  for  $G$ in the  past.  Given  a
typical age of the Universe  $\tau_{U}  \sim\,14$~Gyr  and  assuming a
constant  rate  of  change  it  is  straightforward   to  derive  from
Eq.~(\ref{delta-aa}) the following estimate
\beq
\dGG \sim +10^{-15}\mbox{~yr}^{-1}     \label{dGG}
\eeq 
for the Kaluza-Klein and  the Einstein-Yang-Mills theories, whereas it
is a factor of $10^2$ larger for the Randall-Sundrum scenario.

\begin{figure}[t]
\centering
\includegraphics[clip, width=11 cm]{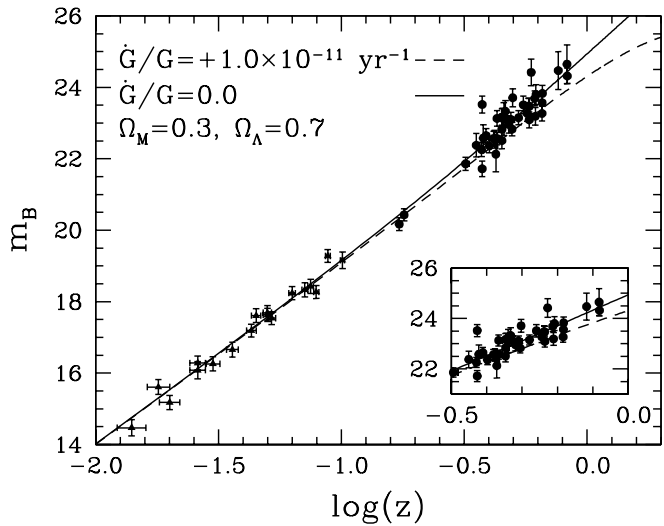}
\caption{The  Hubble  diagram  of  distant  supernovae,  assuming  the
	preferred  cosmological  scenario  of  the SCP,  for  $\dGG=0$
	(solid line), $\dGG  = +10^{-11}$~yr$^{-1}$ (dashed line). For
	$\dGG= +10^{-15}$~yr$^{-1}$  the results are indistinguishable
	from those  of the case with constant  $G$.  The observational
	data are  taken from \cite{SCP}.  The inset  shows an enlarged
	view  of the  region around  $z\sim  0.5$.  See  the text  for
	further details.}
\label{fig1}
\end{figure}

Let  us  see  if  this  estimate   matches  the  currently   available
experimental  bounds on the variation of $G$.  Several  constraints on
the {\sl local} rate of change of $G$ using the  observation  of lunar
occultations   and  eclipses,   planetary   and  lunar   radar-ranging
measurements, the evolution of the Sun, gravitational  lensing, Viking
landers or data from the binary pulsar PSR 1913+16 have been  obtained
up  to  now.  First,  it  is  important  to  realize  that  all  these
measurements  give just upper bounds on the rate of variation  of $G$.
Among these  measurements,  the last one  provided  for many years the
most reliable upper bound \cite{Dam88}
\[ 
-(1.10\pm  1.07) \times 10^{-11}  \mbox{yr}^{-1} < \dGG < 0.  
\] 
However,   the   best   upper   bound   has   been   obtained    using
helioseismological data \cite{Gue98}:
\[ 
-1.6 \times  10^{-12}  \mbox{yr}^{-1}  <  \dot{G}/G  < 0.  
\] 
Note that all these  upper  bounds  are  local.  At {\sl  cosmological
distances} the best upper bound for the rate of variation of $G$ comes
from the Hubble diagram of distant type Ia supernovae.  By taking into
account that distant Type Ia supernovae appear to be {\sl dimmer} than
local supernovae  \cite{SCP,  HzSST}, the following upper bound on the
variation of $G$ was obtained \cite{Gazt01}:

\begin{figure}[t]
\centering
\includegraphics[clip, width=11 cm]{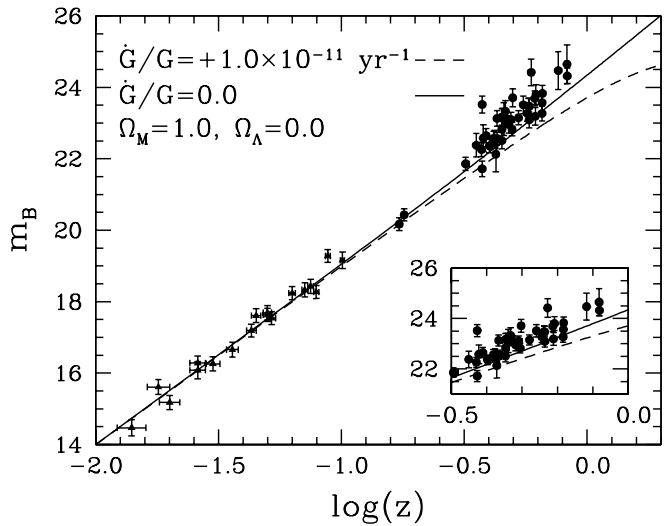}
\caption{Same  as in  Fig.~1,  but assuming  a flat,  matter-dominated
	universe, $(\Omega_M,\Omega_\Lambda)=(1.0,0.0)$.}
\label{fig2}
\end{figure}

\[
-10^{-11} \;\mbox{yr}^{-1}\lapprox \dGG < 0 \; \; \; 
\mbox{at} \;\;\; z \simeq 0.5.  
\]
We would like to stress that  all these upper bounds --- regardless if
they  are local  or  obtained  at moderately  high  redshifts ---  are
negative and, consequently, positive  values of $\dot{G}/G$ seem to be
not allowed  by the  present astrophysical data.   Consequently, these
upper  bounds  are then  {\sl  at odds}  with  the  estimate given  by
Eq.~(\ref{dGG}) which  was obtained  from the recent  determination of
$\daa$,  Eq.~(\ref{delta-aa}),   within  the  multidimensional  models
studied in  Sect.~\ref{KKmodels}.  In the following  we will elaborate
on this.  It  is as well worth mentioning at this  point that the Oklo
natural  nuclear  reactor  \cite{Fujii,  Oklo} severely  constrains  a
variation of $\alpha$ but, again, at low $z$.

Fig.~\ref{fig1} shows the Hubble diagram of distant Type Ia supernovae
for three  values of  $\dGG$  and for the  preferred  scenario  of the
Supernova   Cosmology   Project,   namely   $(\Omega_{R},    \Omega_M,
\Omega_\Lambda)  = (0.0, 0.3,  0.7)$.  For the sake of  simplicity  we
have assumed that $\dGG$  remains  constant  within this $z$ interval.
As it can be seen, the effect of a {\sl  positive}  value of $\dGG$ is
to make distant supernovae to appear {\sl brighter}.  This behavior is
just the opposite to what it is observationally found.  However, as it
can be seen in Fig.~\ref{fig1},  the effect of a varying $G$ at a rate
given by Eq.~(\ref{dGG}), which is the value derived from the observed
variation of $\alpha$, is practically  indistinguishable  from that of
$\dGG=0$.  For the purpose of illustration we also show the curve with
$\dGG=+10^{-11}$~yr$^{-1}$.  Note  also  that  the  case   $\dGG\simeq
+10^{-13}$~yr$^{-1}$  predicted in the Randall-Sundrum-type  model ---
see Eq.~(\ref{beta-def})  --- is bracketed by these two curves.  There
are other  scenarios  which  have been  generally  used to  explain an
accelerating  universe  without the need of  invoking a  non-vanishing
cosmological  constant by attributing this behavior to the presence of
extra  dimensions.  To this  regard  in  Fig.~\ref{fig2}  we show  the
Hubble  diagram for a flat  matter  dominated  universe  $(\Omega_{R},
\Omega_{M},  \Omega_{\Lambda})  = (0.0, 1.0, 0.0)$.  As it can be seen
in  this  figure   there  is  no  possible   way  to   reconcile   the
multidimensional  models considered in Sect.  \ref{KKmodels}  with the
observations no matter which the value of $\dGG \geq 0$ is.

The conclusions above can be put into a quantitative  form.  Using the
$2\sigma$ confidence contours for $z=0.5$ obtained from the fit to the
Hubble diagram of Type Ia supernovae  given in  \cite{Gazt01}  we have
calculated bounds to $\dGG$.  For the currently  favored  cosmological
scenario $(\Omega_{R}, \Omega_{M},  \Omega_{\Lambda}) = (0.0,0.3,0.7)$
and for the  flat  matter  dominated  case  $(\Omega_{R},  \Omega_{M},
\Omega_{\Lambda})=(0.0,1.0,0.0)$   we  have   obtained  the  following
estimates respectively:

\medskip
\begin{tabular}{ll}
 $\Omega_{R}=0.0, \; \Omega_{M}=0.3, \; 
\Omega_{\Lambda}=0.7 \; \; \;$ & 
$-1.4 \cdot 10^{-11} \; \mbox{yr}^{-1} < \dot G/G < 
 +2.6 \cdot 10^{-11} \; \mbox{yr}^{-1}$,  \\  
 $\Omega_{R}=0.0, \; \Omega_{M}=1.0, \; 
\Omega_{\Lambda}=0.0 \; \; \;$ & 
$-2.9 \cdot 10^{-11} \; \mbox{yr}^{-1} < \dot G/G < 
 -0.3 \cdot 10^{-11} \; \mbox{yr}^{-1}$.  \\
\end{tabular}
\medskip 

\noindent  Here, as  before, we  used  the typical  value $\tau_{U}  =
H_{0}^{-1} = 14  \;$Gyr and assumed a constant rate  of change of $G$.
As  it has been  already discussed  above, the  scenario $(\Omega_{R},
\Omega_{M}, \Omega_{\Lambda})=(0.0,1.0,0.0)$ does  not seem to allow a
positive  $\dGG$.    The  cosmological  scenario   with  $(\Omega_{R},
\Omega_{M},  \Omega_{\Lambda})=(0.0,0.3,0.7)$, which is  the preferred
scenario of the SCP, is among  the allowed ones.  It can be shown that
for  a flat Universe  positive values  of $\dGG$  are allowed  only if
$\Omega_{\Lambda}  \gapprox 0.15$.  Another  interesting case  is, for
instance,  $(\Omega_{R}, \Omega_{M},  \Omega_{\Lambda})  = (0.5,  0.5,
0.0)$, for which we obtain

\[
-2.3 \cdot 10^{-11} \; \mbox{yr}^{-1} < \dot G/G < 
+0.3 \cdot 10^{-11} \; \mbox{yr}^{-1}.  
\]
at the $2\sigma$ confidence level. 
 
\section{Conclusions and caveats}
\label{conclusions}

We  have  derived   the  formulae  for  the  time   variation  of  the
gravitational ``constant'' $G$ and  of the fine structure ``constant''
$\alpha$ for three  classes of models with extra  dimensions.  We have
found  that  such variations  are  related  and  we have  derived  the
explicit relation  --- Eqs.  (\ref{daG-gen})  and (\ref{beta-def}) ---
which does not rely on the specific form of the time dependence of the
scale factor  of extra dimensions  $R(t)$.  For the classes  of models
considered in  \S \ref{KKmodels} the relative sign  between $\daa$ and
$\dGG$ turns  out to  be {\sl model  independent}.  Then,  using these
expressions  and  the reported  variation  of  $\alpha$  based on  the
available  data  obtained  from  distant QSOs  \cite{Mur00},  we  have
derived the estimate $\dGG\sim+10^{-15}$~yr$^{-1}$.  This value of the
time variation of the  gravitational constant makes distant supernovae
to appear {\sl brighter}, in contrast with observations.  However, the
effect is too small to safely discard the classes of models with extra
dimensions  considered here.  In  fact, a  positive rate  of variation
$\dGG  \sim  +1 \cdot  10^{-11  \pm  1}  \; \mbox{yr}^{-1}$  has  been
predicted  within  a   $N=1$  ten-dimensional  supergravity  and  with
non-dynamical  dilaton \cite{WuWa86}.   However,  and as  it has  been
shown  in  \cite{Mae88}, when  the  dilaton  dynamics  are taken  into
account the  rate of change of  $\alpha$ is too small  to be observed.
Let  us mention  that in  this case  the relation  between  $\daa$ and
$\dGG$ appears to be given by Eq.  (\ref{daG-1}).

We  have  also  computed  the  Hubble  diagrams  of  distant  Type  Ia
supernovae  in  the framework  of  these  models  for several  typical
cosmological  scenarios  and   analyzed  their  consistency  with  the
available observational data.  We have  found that, if a flat Universe
is assumed, models with extra  dimensions and {\sl positive} values of
$\dGG$,  in accordance  with Eq.~(\ref{delta-aa}),  can  reproduce the
observational data only if  $\Omega_\Lambda \gapprox 0.15$. One should
however keep  in mind that provided  that the value  of $\daa$ derived
from  QSOs turns  out to  be a  genuine detection  and given  that the
corresponding estimate  of $\dGG$ is  very small, at the  redshifts of
interest  $(z \sim  1.5)$ the  deviation  with respect  to the  Hubble
diagram  for $\dGG=0$  is also  very small  so the  observations still
leave much room for  the multidimensional models studied here provided
that  $\Omega_\Lambda \gapprox  0.15$.  Conversely, for  the class  of
models studied  here to  be able to  reproduce the  observational data
with $\Omega_\Lambda  \lapprox 0.15$  {\sl negative} values  of $\dGG$
are  needed (see  Fig.  2).   However,  the absolute  value of  $\dGG$
needed  to  fit the  observations  should  be  much larger  than  that
obtained here from distant QSOs.

A  natural question  which arises  is  whether the  robustness of  the
observational results  on $\daa$ and  $\dGG$ allows to  safely discard
the multidimensional approach for  scenarios involving a flat Universe
with $\Omega_{\Lambda} \lapprox  0.15$, attending {\sl exclusively} to
the  relative   signs  of   $\daa$  and  $\dGG$.   There  are   a  few
possibilities.   One of  them is  that the  observational data  is not
precise enough at  the moment and does not allow  to draw any definite
conclusions.   This   is  perhaps   the  simplest  and   most  obvious
explanation.  The  situation may improve when, for  example, better or
additional observational determinations of $\daa$ are obtained or when
experimental  data from  distant  supernovae become  available in  the
future  (with missions  like SNAP)  to  give more  accurate bounds  on
$\dGG$.  Although this is indeed the most straightforward explanation,
let us  consider other possibilities.   Firstly, it may turn  out that
our consideration  of the time variation of  the fundamental couplings
is  too rough and  some effects  which modify  Eq.~(\ref{daG-gen}) are
missing.  And  secondly, it could be that  the multidimensional models
considered  here  may  also   be  too  simple  and  phenomenologically
unsatisfactory and should be  substituted by more elaborated ones.  In
any  case,  questioning  the  applicability  of  the  multidimensional
approach for the description of the fundamental interactions should be
preceded by additional more detailed theoretical studies and should be
confronted with more accurate observational data.

\ack We thank E.  Gazta\~naga  for a careful reading of the manuscript
and useful remarks.   We are also grateful to  J.  Garriga, P.  Kanti,
R.  Lehnert, J.  Uzan and  E.  Verdaguer for comments and suggestions.
The   work    of   Yu.~Kubyshin    was   supported   by    the   grant
CERN/FIS/43666/2001 and  RFBR grant  02-02-16444.  This work  has been
partially supported  by the MCYT grant  AYA2002-04094-C03-01/02 and by
the CIRIT.

\end{document}